\documentclass{article}

\usepackage{fullpage}
\usepackage{url}
\usepackage{ifthen}
\usepackage{cite}
\usepackage[cmex10]{amsmath} % Use the [cmex10] option to ensure complicance
\usepackage{amsmath,amsthm,mathtools}
\usepackage{amssymb}
\usepackage{mathrsfs}
\usepackage{algorithmic}

\usepackage{multirow}
\usepackage{array}

\usepackage{amsfonts}
\usepackage{graphicx}
\usepackage[linesnumbered]{algorithm2e}

\usepackage{color, verbatim}
\usepackage[usenames,dvipsnames]{xcolor}
\usepackage{cleveref}
\usepackage[utf8]{inputenc} % allow utf-8 input
\usepackage[T1]{fontenc}    % use 8-bit T1 fonts
\usepackage{url}            % simple URL typesetting
\usepackage{booktabs}       % professional-quality tables
\usepackage{amsfonts}       % blackboard math symbols
\usepackage{nicefrac}       % compact symbols for 1/2, etc.
\usepackage{microtype}      % microtypography
\usepackage{enumitem}

\usepackage{epsfig}
\usepackage{subcaption}
\usepackage{epstopdf}
\usepackage{xargs}
\usepackage{hhline}

\usepackage{threeparttable}
\usepackage{soul}

\newcommand\nc\newcommand
\nc{\bb}[1]{\mathbb{#1}}
\renewcommand{\cal}[1]{\mathcal{#1}}
\renewcommand{\bf}[1]{\mathbf{#1}}

\newcommand{\K}{\mathcal{K}}

\newcommand{\poly}{\textrm{poly}}

\DeclarePairedDelimiter{\set}{\lbrace}{\rbrace}
\DeclarePairedDelimiter{\br}{\lparen}{\rparen}

\Crefname{figure}{Fig.}{Figs.}% {<type>}{<singular>}{<plural>}
\crefname{pluralequation}{eqs.}{eqs.}
\Crefname{pluralequation}{Eqs.}{Eqs.}
\crefformat{pluralequation}{#2(#1)#3}
\Crefformat{pluralequation}{#2(#1)#3}
\crefname{algocf}{algorithm}{algorithms}
\crefname{definition}{definition}{definitions}
\Crefname{algocf}{Algorithm}{Algorithms}
\crefname{definition}{defn.}{defns}
\Crefname{definition}{Definition}{Definitions}

\nc\bfa{{\bf{a}}}\nc\bfA{{\boldsymbol A}}\nc\cA{{\cal A}} \nc\fA[1]{A\br*{#1}} \nc\fa[1]{a\br*{#1}}  \nc\rmA{\mathrm{A}} \nc\rma{\mathrm{a}}
\nc\bfb{{\bf{b}}}\nc\bfB{{\boldsymbol B}}\nc\cB{{\cal B}} \nc\fB[1]{B\br*{#1}} \nc\fb[1]{b\br*{#1}}  \nc\rmB{\mathrm{B}} \nc\rmb{\mathrm{b}}
\nc\bfc{{\bf{c}}}\nc\bfC{{\boldsymbol C}}\nc\cC{{\cal C}} \nc\fC[1]{C\br*{#1}} \nc\fc[1]{c\br*{#1}}  \nc\rmC{\mathrm{C}} \nc\rmc{\mathrm{c}}
\nc\bfd{{\bf{d}}}\nc\bfD{{\boldsymbol D}}\nc\cD{{\cal D}} \nc\fD[1]{D\br*{#1}} \nc\fd[1]{d\br*{#1}}  \nc\rmD{\mathrm{D}} \nc\rmd{\mathrm{d}}
\nc\bfe{{\bf{e}}}\nc\bfE{{\boldsymbol E}}\nc\cE{{\cal E}} \nc\fE[1]{E\br*{#1}} \nc\fe[1]{e\br*{#1}}  \nc\rmE{\mathrm{E}} \nc\rme{\mathrm{e}}
\nc\bff{{\bf{f}}}\nc\bfF{{\boldsymbol F}}\nc\cF{{\cal F}} \nc\fF[1]{F\br*{#1}} \nc\ff[1]{f\br*{#1}}  \nc\rmF{\mathrm{F}} \nc\rmf{\mathrm{f}}
\nc\bfg{{\bf{g}}}\nc\bfG{{\boldsymbol G}}\nc\cG{{\cal G}} \nc\fG[1]{G\br*{#1}} \nc\fg[1]{g\br*{#1}}  \nc\rmG{\mathrm{G}} \nc\rmg{\mathrm{g}}
\nc\bfh{{\bf{h}}}\nc\bfH{{\boldsymbol H}}\nc\cH{{\cal H}} \nc\fH[1]{H\br*{#1}} \nc\fh[1]{h\br*{#1}}  \nc\rmH{\mathrm{H}} \nc\rmh{\mathrm{h}}
\nc\bfi{{\bf{i}}}\nc\bfI{{\boldsymbol I}}\nc\cI{{\cal I}} \nc\fI[1]{I\br*{#1}} \nc\rmI{\mathrm{I}} \nc\rmi{\mathrm{i}}
\nc\bfj{{\bf{j}}}\nc\bfJ{{\boldsymbol J}}\nc\cJ{{\cal J}} \nc\fJ[1]{J\br*{#1}} \nc\fj[1]{j\br*{#1}} \nc\rmJ{\mathrm{J}} \nc\rmj{\mathrm{j}}
\nc\bfk{{\bf{k}}}\nc\bfK{{\boldsymbol K}}\nc\cK{{\cal K}} \nc\fK[1]{K\br*{#1}} \nc\fk[1]{k\br*{#1}} \nc\rmK{\mathrm{K}} \nc\rmk{\mathrm{k}}
\nc\bfl{{\bf{l}}}\nc\bfL{{\boldsymbol L}}\nc\cL{{\cal L}} \nc\fL[1]{L\br*{#1}} \nc\fl[1]{l\br*{#1}} \nc\rmL{\mathrm{L}} \nc\rml{\mathrm{l}}
\nc\bfm{{\bf{m}}}\nc\bfM{{\boldsymbol M}}\nc\cM{{\cal M}} \nc\fM[1]{M\br*{#1}} \nc\fm[1]{m\br*{#1}} \nc\rmM{\mathrm{M}} \nc\rmm{\mathrm{m}}
\nc\bfn{{\bf{n}}}\nc\bfN{{\boldsymbol N}}\nc\cN{{\cal N}} \nc\fN[1]{N\br*{#1}} \nc\fn[1]{n\br*{#1}} \nc\rmN{\mathrm{N}} \nc\rmn{\mathrm{n}}
\nc\bfo{{\bf{o}}}\nc\bfO{{\boldsymbol O}}\nc\cO{{\cal O}} \nc\fO[1]{O\br*{#1}} \nc\fo[1]{o\br*{#1}} \nc\rmO{\mathrm{O}} \nc\rmo{\mathrm{o}}
\nc\bfp{{\bf{p}}}\nc\bfP{{\boldsymbol P}}\nc\cP{{\cal P}} \nc\fP[1]{P\br*{#1}} \nc\fp[1]{p\br*{#1}} \nc\rmP{\mathrm{P}} \nc\rmp{\mathrm{p}}
\nc\bfq{{\bf{q}}}\nc\bfQ{{\boldsymbol Q}}\nc\cQ{{\cal Q}} \nc\fQ[1]{Q\br*{#1}} \nc\fq[1]{q\br*{#1}} \nc\rmQ{\mathrm{Q}} \nc\rmq{\mathrm{q}}
\nc\bfr{{\bf{r}}}\nc\bfR{{\boldsymbol R}}\nc\cR{{\cal R}} \nc\fR[1]{R\br*{#1}} \nc\fr[1]{r\br*{#1}} \nc\rmR{\mathrm{R}} \nc\rmr{\mathrm{r}}
\nc\bfs{{\bf{s}}}\nc\bfS{{\boldsymbol S}}\nc\cS{{\cal S}} \nc\fS[1]{S\br*{#1}} \nc\fs[1]{s\br*{#1}} \nc\rmS{\mathrm{S}} \nc\rms{\mathrm{s}}
\nc\bft{{\bf{t}}}\nc\bfT{{\boldsymbol T}}\nc\cT{{\cal T}} \nc\fT[1]{T\br*{#1}} \nc\ft[1]{t\br*{#1}} \nc\rmT{\mathrm{T}} \nc\rmt{\mathrm{t}}
\nc\bfu{{\bf{u}}}\nc\bfU{{\boldsymbol U}}\nc\cU{{\cal U}} \nc\fU[1]{U\br*{#1}} \nc\fu[1]{u\br*{#1}} \nc\rmU{\mathrm{U}} \nc\rmu{\mathrm{u}}
\nc\bfv{{\bf{v}}}\nc\bfV{{\boldsymbol V}}\nc\cV{{\cal V}} \nc\fV[1]{V\br*{#1}} \nc\fv[1]{v\br*{#1}} \nc\rmV{\mathrm{V}} \nc\rmv{\mathrm{v}}
\nc\bfw{{\bf{w}}}\nc\bfW{{\boldsymbol W}}\nc\cW{{\cal W}} \nc\fW[1]{W\br*{#1}} \nc\fw[1]{w\br*{#1}} \nc\rmW{\mathrm{W}} \nc\rmw{\mathrm{w}}
\nc\bfx{{\bf{x}}}\nc\bfX{{\boldsymbol X}}\nc\cX{{\cal X}} \nc\fX[1]{X\br*{#1}} \nc\fx[1]{x\br*{#1}} \nc\rmX{\mathrm{X}} \nc\rmx{\mathrm{x}}
\nc\bfy{{\bf{y}}}\nc\bfY{{\boldsymbol Y}}\nc\cY{{\cal Y}} \nc\fY[1]{Y\br*{#1}} \nc\fy[1]{y\br*{#1}} \nc\rmY{\mathrm{Y}} \nc\rmy{\mathrm{y}}
\nc\bfz{{\bf{z}}}\nc\bfZ{{\boldsymbol Z}}\nc\cZ{{\cal Z}} \nc\fZ[1]{Z\br*{#1}} \nc\fz[1]{z\br*{#1}} \nc\rmZ{\mathrm{Z}} \nc\rmz{\mathrm{z}}

\nc\defeq{\coloneqq}

\DeclarePairedDelimiterX\Set[1]\{\}{#1}

\newcommand{\E}{\mathbb{E}}

\newtheorem{theorem}{Theorem}

\newtheorem{lemma}[theorem]{Lemma}
\newtheorem*{lemma*}{Lemma}

\newtheorem*{corollary*}{Corollary}

\newtheorem{remark*}{\indent Remark}

\begin{document}
\title{Probabilistic Group Testing with a Linear Number of Tests} 

% %%% Single author, or several authors with same affiliation:
% \author{%
%   \IEEEauthorblockN{Stefan M.~Moser}
%   \IEEEauthorblockA{ETH Zürich\\
%                     ISI (D-ITET)\\
%                     CH-8092 Zürich, Switzerland\\
%                     Email: moser@isi.ee.ethz.ch}
% }

%%% Several authors with up to three affiliations:
\author{%
  Larkin Flodin \\
  %College of Information and Computer Sciences\\
  University of Massachusetts, Amherst\\
  Amherst, MA, 01003 \\
  \texttt{lflodin@cs.umass.edu} \\
  \and
  Arya Mazumdar \\
  University of California, San Diego\\
    La Jolla, CA, 92093\\
    \texttt{arya@ucsd.edu}
}
\date{}

%%% Many authors with many affiliations:
% \author{%
%   \IEEEauthorblockN{Albus Dumbledore\IEEEauthorrefmark{1},
%                     Olympe Maxime\IEEEauthorrefmark{2},
%                     Stefan M.~Moser\IEEEauthorrefmark{3}\IEEEauthorrefmark{4},
%                     and Harry Potter\IEEEauthorrefmark{1}}
%   \IEEEauthorblockA{\IEEEauthorrefmark{1}%
%                     Hogwarts School of Witchcraft and Wizardry,
%                     1714 Hogsmeade, Scotland,
%                     \{dumbledore, potter\}@hogwarts.edu}
%   \IEEEauthorblockA{\IEEEauthorrefmark{2}%
%                     Beauxbatons Academy of Magic,
%                     1290 Pyrénées, France,
%                     maxime@beauxbatons.edu}
%   \IEEEauthorblockA{\IEEEauthorrefmark{3}%
%                     ETH Zürich, ISI (D-ITET), ETH Zentrum, 
%                     CH-8092 Zürich, Switzerland,
%                     moser@isi.ee.ethz.ch}
%   \IEEEauthorblockA{\IEEEauthorrefmark{4}%
%                     National Chiao Tung University (NCTU), 
%                     Hsinchu, Taiwan,
%                     moser@isi.ee.ethz.ch}
% }

\maketitle

{\renewcommand{\thefootnote}{}\footnotetext{
This work was supported in part by NSF CCF Awards 1526763, 1909046, and 1934846.
}
\renewcommand{\thefootnote}{\arabic{footnote}}
\setcounter{footnote}{0}

%%%%%%
%% Abstract: 
%% If your paper is eligible for the student paper award, please add
%% the comment "THIS PAPER IS ELIGIBLE FOR THE STUDENT PAPER
%% AWARD." as a first line in the abstract. 
%% For the final version of the accepted paper, please do not forget
%% to remove this comment!
%%
\begin{abstract}
   In probabilistic nonadaptive group testing (PGT), we aim to characterize the number of pooled tests necessary to identify a random $k$-sparse vector of defectives with high probability. Recent work has shown that $n$ tests are necessary  when $k =\omega(n/\log n)$. It is also known that $O(k \log n)$ tests are necessary and sufficient in other regimes. This leaves open the important sparsity regime where the probability of a defective item is $\sim 1/\log n$ (or $k = \Theta(n/\log n)$) where the number of tests required is linear in $n$.  In this work we aim to exactly characterize the number of tests in  this sparsity regime. In particular, we seek to determine the number of defectives $\lambda(\alpha)n / \log n$ that can be identified if the number of tests is $\alpha n$. In the process, we give upper and lower bounds on the exact point at which individual testing becomes suboptimal, and the use of a carefully constructed pooled test design is beneficial.
\end{abstract}

%% The paper must be self-contained. However, if you are referring to
%% a full version for checking certain proofs, please provide the
%% publically accessible location below.  If the paper is completely
%% self-contained, you can remove the following line from your
%% submission.
% \textit{A full version of this paper is accessible at:}
% \url{https://arxiv.org/pdf/21xx.xxxx.pdf} 

\section{Introduction}

Group testing is a sparse recovery problem where we aim to recover a small set of $k$ ``defective'' items from among $n$ total items using pooled tests. Originally introduced in the context of testing blood samples for diseases where multiple samples can be combined together \cite{dorfman1943detection}, it has since seen a variety of applications, including recently COVID-19 testing (see for instance \cite{yelin2020evaluation, 
%petersen2020practical, mutesa2020strategy,
gollier2020group, eberhardt2020multi}, though there are many more such works).

More formally, we let $\bfx \in \set{0, 1}^n$ represent the vector of items where a 1 indicates defectivity. A test vector $\bfm \in \set{0, 1}^n$ specifies a subset of items to be tested; the result is
\begin{equation*}
    y = \bigvee_{i:\bfm_i = 1} \bfx_i,
\end{equation*}
the logical OR of the entries of $\bfx$ specified by the test.

We focus here on the nonadaptive setting, where all tests must be specified in advance of seeing any results. In this setting we write $M \in \set{0,1}^{T \times n}$ for the matrix of $T$ tests (rows), and $\bfy$ for the vector of all test results. We assume the vector of defectives $\bfx$ is generated randomly, the details of which are discussed in \cref{sec:priors}.

This is known as \emph{probabilistic group testing} (PGT), meaning that we are interested in determining how many tests are necessary for the probability of error (for some particular decoding method) to approach 0 as $n$ becomes large. This is in contrast to \emph{combinatorial group testing}, where we insist that any set of defectives is recoverable (i.e., probability of error is 0 for any $n$).

In general, we are interested in the question of how the number of tests $T$ must scale as a function of both $n$ and $k$. This behavior can be quite different depending on the scaling of $k$ relative to $n$, so often things are broken down further into different ``regimes'' for $k$. We will in this work prove both lower and upper bounds on $T$ for the specific regime $k = \Theta\left(\frac{n}{\log n}\right)$, or equivalently when the probability of an item being defective is $\Theta\left(\frac1{\log n}\right)$; this regime has traditionally seen little attention, but recent work has shown that this is exactly the scaling of $k$ for which $\Omega(n)$ tests are necessary. As $n$ tests trivially suffice by testing each item individually, determining the exact crossover point where individual testing becomes suboptimal seems of interest, which happens to be in  the $k = \Theta\left(\frac{n}{\log n}\right)$ regime.

\subsection{Related Work}
\label{sec:related_work}

Much of the initial work on group testing concerned the zero-error combinatorial setting, where a single test matrix must correctly classify every possible defective vector. The literature on this variant is vast -- see for instance the book of Du and Hwang \cite{du2000combinatorial}. We focus the rest of this section on the low-error probabilistic setting.

PGT has traditionally been split into two rather different regimes: the ``sparse'' regime, where the total number of defectives is $O(n^\theta)$ for some $0 \leq \theta < 1$, and the ``linear'' regime where the total number of defectives is $\beta n$ for some constant $\beta < 1$.

In both regimes, the folklore ``counting bound'' (see \cite{chan2011non} for instance) shows that $\Omega(k \log \frac{n}{k})$ measurements are necessary. In the sparse regime, this is equivalent to $\Omega(k \log n)$, and order-optimal randomized constructions have been known for some time \cite{sebHo1985two, atia2012boolean}. In the linear regime, the counting bound implies $\Omega(n)$ measurements are necessary, and trivially $n$ suffice by testing items individually.

More recently, explicit constructions of matrices for PGT have been studied. Mazumdar \cite{mazumdar2015nonadaptive} gave explicit constructions requiring $O(k \log^2 n / \log k)$ measurements. Follow up works of Barg and Mazumdar~\cite{barg2017group}, and Inan et al. \cite{inan2019optimality},  show order-optimal or near-optimal results. % $O(k \log n)$ measurements when $k = \Omega(\log ^2 n)$.

Another direction has been to develop constructions with good decoding properties. PGT schemes are considered efficiently decodable if they require $O(T)$ time to decode, where $T$ is the total number of measurements; several works \cite{cai2017efficient, vem2017group, lee2019saffron} gave efficiently decodable schemes which were not quite order-optimal, before Bondorf et al. \cite{bondorf2020sublinear} gave the first order-optimal efficiently decodable construction. A very recent result \cite{inan2020strongly} shows that we can even have explicit constructions which are both order-optimal and efficiently decodable.

The last line of work we discuss in PGT, and the most relevant to our work here, has been to go beyond order-optimality and determine the precise constants involved in various regimes.  Improvements on the counting bound were proposed in \cite{agarwal2018novel} and subsequently it was shown the individual testing is optimal in  the linear regime~\cite{aldridge2018individual}.

In the sparse regime the characterization of exact constants results in a more complex picture, but a series of works \cite{scarlett2016limits, aldridge2017capacity, johnson2018performance, coja2020optimal} have narrowed down the constants to the point that the lower and upper bounds are matching for any $\theta \in [0, 1)$ (where $k = n^\theta$). We direct the interested reader to the recent survey of Aldridge et al. \cite{aldridge2019group}.

However, in between these two regimes is another regime which has seen much less study, namely when the total number of defectives is $n / \poly(\log n)$ (\hspace{1sp}\cite{bay2020optimal} refers to these regimes as ``mildly sublinear''). In particular when $k = \Theta(n / \log n)$ the counting bound implies only that $\Omega(n \log \log n/ (\log n))$ measurements are necessary, but a method using $o(n)$ measurements proved elusive. In very recent work, Bay et al. \cite{bay2020optimal} showed that in fact the lower bound can be improved to $\Omega(k \log n)$ in this regime, and thus known constructions are order-optimal.

In light of the asymptotic improvement to the lower bound for $k = n / \poly(\log n)$ \cite{bay2020optimal}, we feel it is a natural next step to try and nail down the constants in this regime. While we focus on the particular case $k = \lambda n / \log n$ in this paper, we expect our results should extend readily to other mildly sublinear regimes.

\subsection{Priors for PGT}
\label{sec:priors}

%\textcolor{red}{This section could be cut down a bit, but our UB uses combinatorial prior and the LB iid prior so some mention is needed.}

There are two commonly used ``priors'' over the random defective vector involved in PGT. Under the \emph{combinatorial prior}, the number of defectives is fixed to be $k$, and the defective set is drawn uniformly from the $\binom{n}{k}$ possibilities. Under the \emph{i.i.d. prior}, we instead fix a defective probability $p$, and each item is defective independently with probability $p$.

Conveniently, it turns out that our choice of prior makes little difference in the number of tests necessary (assuming we take $k = pn$). The following result from the survey of Aldridge et al. \cite{aldridge2019group} formalizes this notion.

\begin{theorem}[\hspace{1sp}\cite{aldridge2019group} Thm. 1.7]
\label{thm:comb_prior_conversion}
Suppose a sequence of test designs and decoding method has probability of error going to 0 as $n$ goes to infinity under the combinatorial prior with $k = k_0(1 + o(1))$, where $k_0 = o(n)$ and $k_0$ goes to infinity with $n$. Then the same sequence of test designs and decoding method has error going to 0 as $n$ goes to infinity under the i.i.d. prior with $p = k_0 / n$.
\end{theorem}
A similar result holds for converting the opposite direction. In this paper we will primarily use the i.i.d. prior, except for our upper bound in \cref{sec:constant_ub} which relies on preexisting work under the combinatorial prior. For ease of reference to that work we will use the combinatorial prior there, and \cref{thm:comb_prior_conversion} tells us we can convert our main result (\cref{thm:constant_ub}) to an equivalent result under the i.i.d. prior.

\subsection{Notation}

We will use the following notational conventions throughout:
\begin{itemize}
    \item $M$ is a test matrix with rows corresponding to tests and columns corresponding to items.
    \item $T$ is the number of tests (rows of $M$).
    \item $n$ is the total number of items (columns of $M$).
    \item $k$ is the total number of defectives.
    \item $p$ is the defective probability under the i.i.d. prior.
    \item $\log$ is the logarithm base $e$. %\textcolor{red}{we could use $\ln$ instead}.
    \item $\epsilon$, $\delta$, $\gamma$, $\lambda$, $\alpha$, $\nu$ are constants.
\end{itemize}

Our  results show that when $p = \lambda / \log n$, it is sufficient to have $T/n \geq \min(1,\frac{\lambda}{\log^2 2})$, whereas $T/n \geq \frac{\lambda}{\lambda+\log^2 2}$ is necessary. Note that these two quantities are always within a factor of 2 of each other, and approach equality for very small or large values of $\lambda$.

\section{Upper Bound with Near-Constant Tests-Per-Item Design}
\label{sec:constant_ub}

For probabilistic group testing in the sparse regime, it has been shown that the optimal rate is achieved by so-called ``Near-Constant Tests-per-Item'' designs \cite{johnson2018performance}. These matrices are formed by drawing $\frac{\nu T}{k}$ tests uniformly with replacement per item, where $\nu$ is a small constant parameter to be optimized, $T$ is the total number of tests, and $k$ is the number of defectives (assuming the combinatorial prior). 

Drawing the tests with replacement means that it is possible that the same test is drawn more than once, hence the ``near-constant'' tests-per-item, rather than constant. This simplifies the analysis as the draws are independent.

We will use this type of test matrix to prove our upper bound.

\subsection{Preliminaries}

We closely follow the approach taken by Johnson et al. \cite{johnson2018performance} for the sparse regime in the following, as their argument is mostly regime-independent. The following notation will be useful in describing their approach succinctly:
\begin{itemize}
    \item $\K$ is the set of all defective items.
    \item $W^{(\K)}$ is the number of tests containing at least one item from $\K$.
    \item $G$ is the number of nondefective items that do not appear in any negative tests.
\end{itemize}

Our upper bound will employ the simple COMP decoding~\cite{du2000combinatorial,chan2011non}. This algorithm works by first identifying all items present in negative tests and classifying them as nondefective. Then, any remaining items are classified as defective. While simple, in most cases this algorithm has proven to be asymptotically as good as more complex decodings, and is easy to analyze.

First we present two useful lemmas borrowed from \cite{johnson2018performance}. The first in plain language states that given a near-constant tests-per-item design, the number of tests including at least one defective concentrates tightly around its mean.

\begin{lemma}[\hspace{1sp}\cite{johnson2018performance} Lemma 1]
\label[lemma]{thm:wk_concentration}
Let $|\mathcal{K}| = k$, and fix constants $\alpha > 0$, $\epsilon \in (0, 1)$. Suppose we form a test design of $T$ tests by drawing $L = \frac{\alpha T}{k}$ tests uniformly at random with replacement for each of $n$ items. Then (assuming $T$ goes to infinity with $n$) the following holds:
\begin{equation*}
    \Pr[|W^{(\mathcal{K})} - (1 - e^{-\alpha})T| \geq \delta] \leq 2 \exp \left( -\frac{\delta^2}{\alpha T} \right).
\end{equation*}
\end{lemma}

The next lemma, again from \cite{johnson2018performance}, characterizes the distribution of $G$ conditioned on a fixed value of $W^{(\mathcal{K})}$.

\begin{lemma}[\hspace{1sp}\cite{johnson2018performance} Lemma 5]
\label[lemma]{thm:g_dist}
Let $L = \frac{\nu T}{k}$ be the number of tests drawn for each item in a near-constant tests-per-item design. Then
\begin{equation*}
    G \ | \ (W^{(\mathcal{K})} = x) \sim \mathrm{Bin} (n-k, (x/T)^L).
\end{equation*}
\end{lemma}

\subsection{Combining Together}

Now we are ready to prove the upper bound, namely that we can succeed with high probability using about $n \cdot \lambda/\log^2 2$ tests.

\begin{theorem}
\label{thm:constant_ub}
Let $k = \frac{\lambda n}{\log n}$ be the total number of defectives. Suppose our test design is near-constant tests-per-item,
\begin{equation*}
    T = (1 + \epsilon) \frac{\lambda}{\log^2 2} n
\end{equation*}
for a constant $\epsilon > 0$, and we draw $L = \frac{T \log 2}{k}$ tests per item with replacement. Then for sufficiently large $n$, the success probability of the COMP decoding is $1 - o(1)$.
\end{theorem}
\begin{proof}
As $G$ is the number of nondefectives not appearing in any negative tests, we know COMP succeeds if and only if $G = 0$. Then the main idea here is that we will use the equation
\begin{equation}
\label{eq:cond_prob_sum}
    \Pr[G = 0] = \sum_{x=0}^T \Pr[W^{(\mathcal{K})} = x] \Pr[G=0 | W^{(\mathcal{K})} = x],
\end{equation}
applying \Cref{thm:g_dist} when $x \leq (1/2 + \delta) T$, and showing that the probability of $x > (1/2 + \delta) T$ goes to 0 for large $n$ using \Cref{thm:wk_concentration}.

By \Cref{thm:g_dist},
\begin{equation*}
\label{eq:success_conditional_exact}
    \Pr[G = 0 \ | \ W^{(\mathcal{K})} = x] = \left( 1 - \left( \frac{x}{T} \right)^L \right)^{n-k}.
\end{equation*}

As this function is decreasing in $x$, for all $x \leq (1/2 + \delta)T$, we have
\begin{equation*}
    \Pr[G = 0 \ | \ W^{(\mathcal{K})} = x] \geq \Pr[G = 0 \ | \ W^{(\mathcal{K})} = (1/2 + \delta)T]
    = (1 - (1/2 + \delta)^L)^{n-k}.
\end{equation*}

By definition
\begin{equation*}
    L = \frac{T \log 2}{k} = \frac{(1+\epsilon) \lambda n \log 2 \log n}{(\log 2)^2 \lambda n} = \frac{(1+\epsilon) \log n}{\log 2},
\end{equation*}
so this gives
\begin{equation*}
    \frac{1}{2^L} = \left( \frac{1}{2^{1 / \log 2}} \right)^{(1+\epsilon) \log n} = \left(\frac{1}{e^{\log n}} \right)^{(1+\epsilon)} = \frac{1}{n^{1+\epsilon}}.
\end{equation*}
Taking $\delta < 2/L$, the maximum term in the binomial expansion of $(1/2 + \delta)^L$ will be $1/2^L$, so we have
\begin{equation*}
    (1/2 + \delta)^L < \frac{L+1}{2^L} = \frac{L+1}{n^{1+\epsilon}} < \frac{1}{n^{1+\epsilon/2}},
\end{equation*}
where in the last step we use the fact that $L = \Theta(\log n)$ and the assumption regarding large $n$. Then we have for $x \leq (1/2 + \delta)T$
% \begin{align*}
%     \Pr[G = 0 \ | \ W^{(\mathcal{K})} = x] >& \left(1 - \frac{1}{n^{1+\epsilon/2}} \right)^{n-k} \\
%     \approx& 1 - \frac{n-k}{n^{1+\epsilon/2}} \\
%     >& 1 - \frac{n}{n^{1+\epsilon/2}} \\
%     =& 1 - \frac{1}{n^{\epsilon/2}},
% \end{align*}
\begin{equation*}
    \Pr[G = 0 \ | \ W^{(\mathcal{K})} = x] > \left(1 - \frac{1}{n^{1+\epsilon/2}} \right)^{n-k}
    \approx 1 - \frac{n-k}{n^{1+\epsilon/2}}
    > 1 - \frac{n}{n^{1+\epsilon/2}}
    = 1 - \frac{1}{n^{\epsilon/2}},
\end{equation*}
which goes to 1 for large $n$.

Plugging this back into \eqref{eq:cond_prob_sum},
\begin{align}
    \Pr[G = 0] =& \sum_{x=0}^T \Pr[W^{(\mathcal{K})} = x] \cdot \Pr[G=0 \ | \ W^{(\mathcal{K})} = x] \nonumber \\
    \geq& \sum_{x=0}^{(1/2 + \delta)T} \Pr[W^{(\mathcal{K})} = x] \cdot \Pr[G=0 \ | \ W^{(\mathcal{K})} = x] \nonumber \\
    >& \sum_{x=0}^{(1/2 + \delta)T} \Pr[W^{(\mathcal{K})} = x] \left( 1 - \frac{1}{n^{\epsilon/2}} \right) \nonumber \\
    =& \left( 1 - \frac{1}{n^{\epsilon/2}} \right) \cdot \Pr[W^{(\mathcal{K})} \leq (1/2 + \delta)T] \nonumber \\
    =& \left( 1 - \frac{1}{n^{\epsilon/2}} \right) \cdot \left( 1 - \Pr[W^{(\mathcal{K})} > (1/2 + \delta)T]\right) \label{eq:final_succ_prob}.
\end{align}

Finally, we apply \Cref{thm:wk_concentration} to the latter probability with $\alpha = \log 2$ (and $\delta T$ in place of the $\delta$ in the original lemma statement), which tells us
\begin{equation*}
   \Pr[|W^{(\mathcal{K})} - (1 - e^{-\log 2})T| \geq \delta T] \leq 2 \exp \left( -\frac{(\delta T)^2}{T \log 2} \right),
\end{equation*} 
and thus
\begin{equation*}
   \Pr[|W^{(\mathcal{K})} - (T/2)| \geq \delta T] \leq 2 \exp \left( -\frac{\delta^2 T}{\log 2} \right)
  = \Theta \left( \exp \left( -\frac{n}{\log^2 n}\right)\right),
\end{equation*}
which clearly goes to 0 for large $n$. Thus the success probability in \eqref{eq:final_succ_prob} will go to 1, concluding the proof.
\end{proof}

\section{Determining the Constant in the Lower Bound}

The recent work of Bay et al. \cite{bay2020optimal} shows that a lower bound of $\min(n, \Omega(k \log n))$ tests holds for any $k$. In the case we are interested in that $k = \lambda n / \log n$, this tells us we need $\Omega(n)$ measurements. In this section we seek to determine the necessary number of measurements more precisely, up to the constant term. 

The argument of \cite{bay2020optimal} works by demonstrating that with significantly less than $k \log n$ measurements, there must exist a fairly large number of ``totally disguised'' items; that is, items for which no test gives us any information about whether or not these items are defective. If we have no information about these items defectivity, we cannot do better than guessing on each one.

More specifically, they give a procedure that constructs a set of items $W$ with the following properties:
\begin{enumerate}
    \item Each item in $W$ is totally disguised with probability at least $\mathcal{L}^*$.
    \item The events that each item in $W$ is totally disguised are independent of each other.
\end{enumerate}

%\textcolor{red}{Do we need to define Procedure 1 here? We are modifying it slightly, and it would make things more self-contained, but might be tricky with the page limit.}

Since these events are independent, we can combine bounds on $|W|$ and $\mathcal{L}^*$ with simple concentration inequalities to get a lower bound on the number of totally disguised items.

For our lower bound we will modify their procedure slightly, as described below.

\subsection{How Many Disguised Items are Needed}
Since disguised items are defective independently with probability $p$ and for us $p < \frac{1}{2}$, the best any algorithm
can do on a disguised item is predict it is nondefective. This guess is correct with probability $1 - p$. Thus if the total number of disguised items is $D$, the success probability of any algorithm is at most
\begin{equation*}
    (1 - p)^D \leq \exp(-p D).
\end{equation*}

\subsection{Lower Bounding $|W|$}

We will closely follow the method of \cite{bay2020optimal}, but will first slightly redefine their notion of ``very-present'' items. These are items present in far more tests than the average item, which will be discarded in a preprocessing step before constructing $W$.

\begin{lemma}[\hspace{1sp}\cite{bay2020optimal} Lemma 4, modified]
\label[lemma]{thm:very_present}
Define an item to be \emph{very-present} if it appears in more than $t_{max} = \log^3 n$ tests. If $T \leq n$ and no test contains more than $z \log n$ items, then the number of very-present items is
\begin{equation*}
    n_{vp} \leq \frac{nz}{\log^2 n}.
\end{equation*}
\end{lemma}
\begin{proof}
Consider the total number of (item, test) pairs $P$. From the assumptions, we have $P \leq Tz \log n \leq nz \log n$. Also, by the definition of very-present items, we have $n_{vp} \log^3 n \leq P$. Then $n_{vp} \log^3 n \leq nz \log n$ implies the result.
\end{proof}

For bounding $|W|$, we have the following:
\begin{lemma}[\hspace{1sp}\cite{bay2020optimal} Lemma 6 Pf.]
\label[lemma]{thm:w_lb}
Suppose $W$ is constructed as described in Procedure 1 of \cite{bay2020optimal}, and furthermore we modify their stopping rule so that we stop when there are less than $\frac{\epsilon n}{1 + \gamma}$ items rather than $\frac{\epsilon n}{2}$, where $\gamma$ is a small positive constant. Then
    \begin{equation}
    \label{eq:w_lb}
        |W| \geq \frac{(\gamma \epsilon n) / (1+\gamma) - (nz)/\log^2 n}{z^2 \log^8 n},
    \end{equation}
where $z = \frac{2}{\log (1/(1-p))}$.
\end{lemma}

\begin{proof}
From the proof of \cite{bay2020optimal} Lemma 6, we have that the method of constructing $W$ yields the inequality
\begin{equation*}
    \frac{\epsilon n}{1 + \gamma} \geq \epsilon n - n_{vp} - |W| t_{max}^2 z^2 \log^2 n,
\end{equation*}
where $n_{vp}$ and $t_{max}$ are as defined in our \Cref{thm:very_present}. Rearranging this yields
\begin{equation*}
    |W| \geq \frac{\gamma \epsilon n/(1+\gamma) - n_{vp}}{t_{max}^2 z^2 \log^2 n},
\end{equation*}
and substituting the values from \Cref{thm:very_present} for $n_{vp}$ and $t_{max}$ gives the result.
\end{proof}

In the case we are interested in that $p = \frac{\lambda}{\log n}$, we have
\begin{equation*}
    z = -2 / \log \left(1 - \frac{\lambda}{\log n}\right) =  \frac{2 \log n}{\lambda} - 1 - o(1),
    %2 (\log \log n - \log (\log n - \lambda)),
\end{equation*}
where we have considered the Taylor series as $n$ goes to infinity.
%which as $n$ goes to infinity has Taylor series
%\begin{equation*}
%    \frac{2 \log n}{\lambda} - 1 - o(1).
%\end{equation*}
Then
\begin{equation*}
\frac{nz}{\log^2 n} \approx \frac{2 n \log n}{\lambda \log^2 n} = \frac{2 n}{\lambda \log n} = o(n),
\end{equation*}
so asymptotically we have
\begin{equation*}
    (\gamma n) / (1+\gamma) - (nz)/\log^2 n = \Theta(n).
\end{equation*}
Substituting the value of $z$ back into the denominator of \eqref{eq:w_lb} gives
\begin{equation*}
    |W| \geq \frac{cn}{\log^{10} n}
\end{equation*}
as $n$ goes to infinity, for some constant $c>0$. This will be sufficient for our purposes, as it will turn out that only the exponent of $n$ is relevant for this term.

\subsection{Lower Bounding $\mathcal{L^*}$}

For lower bounding $\mathcal{L}^*$, they show in \cite{bay2020optimal} that if we stop constructing $W$ when there are less than $n_{final}$ items remaining, then
\begin{equation*}
    \mathcal{L}^* \geq \exp \left( \frac{T}{n_{final}} \mathcal{L}_p\right),
\end{equation*}
where
\begin{equation*}
    \mathcal{L}_p = \min_{x=2, 3, \dotsc, n} x \log (1 - (1-p)^{x-1}).
\end{equation*}

As we modified their Procedure 1 to stop with less than $\frac{\epsilon n}{1 + \gamma}$ items, we will have
\begin{equation}
\label{eq:our_lstar}
    \mathcal{L}^* \geq \exp \left( \frac{(1+\gamma)T}{\epsilon n} \mathcal{L}_p\right).
\end{equation}

Furthermore, it is shown in work of Coja-Oghlan et al. \cite{coja2020optimal} in the sparse regime that the function $\mathcal{L}_p$ is minimized at
\begin{equation*}
    x = \frac{\log 2}{p} + O(p^{-1/2}).
\end{equation*}

In our regime neglecting lower order terms, this yields
\begin{equation*}
    \mathcal{L}_p \geq \frac{\log n \log 2}{\lambda} \log \left( 1 - \left(\frac{\log n - \lambda}{\log n} \right)^{(\log n \log 2)/\lambda - 1}\right).
\end{equation*}

From this, we have
\begin{equation*}
    \exp(\mathcal{L}_p) \geq \left( 1 - \left( \frac{\log n - \lambda}{\log n} \right)^{\log 2 \log n / \lambda - 1}\right)^{\log 2 \log n / \lambda}.
\end{equation*}
The Taylor series expansion at $x = \infty$ of
\begin{equation*}
    \left( \frac{x-\lambda}{x} \right)^{\log 2 x/\lambda - 1}
\end{equation*}
is $\frac{1}{2} + O(\frac{1}{x})$, so for large $n$ dropping lower order terms this gives
\begin{equation*}
    \exp(\mathcal{L}_p) \geq \left( \frac{1}{2} \right)^{\log 2 \log n / \lambda},
\end{equation*}
and thus substituting $T = (1-\epsilon)n$ into our expression for $\mathcal{L}^*$ from \eqref{eq:our_lstar},
\begin{equation*}
    \mathcal{L}^* \geq \left( \frac{1}{2} \right)^{(\log 2 \log n / \lambda) \cdot (1+\gamma)(1 - \epsilon)/\epsilon},
\end{equation*}
which can be simplified to
\begin{equation*}
    \mathcal{L}^* \geq \left( \frac{1}{n} \right)^{(1+\gamma)(\log 2)^2 (1 - \epsilon)/(\lambda \epsilon)}.
\end{equation*}

%\textcolor{red}{We could be more rigorous here about handling the lower order terms, but it will make the equations a lot messier.}
    
\subsection{Combining Together}
Writing $D$ for the total number of disguised items, we have $\E[D] \geq \mathcal{L}^* |W|$, and as each item in $W$ is disguised independently, by a Chernoff bound we have $D \geq (1 - o(1)) \mathcal{L}^* |W|$ with high probability. Any algorithm's success probability is upper bounded by $\exp(-p D)$, so in order for this success probability to go to 1, we need $pD$ to go to 0. Then substituting in our bounds for $\mathcal{L}^*$ and $|W|$, we need
\begin{align*}
    -pD \leq& -p \mathcal{L}^* |W| \\
    \leq& \frac{-\lambda}{\log n} \left( \frac{1}{n} \right)^{(1+\gamma)(\log 2)^2(1 - \epsilon)/(\lambda \epsilon)} \frac{cn}{\log^{10} n} \\
    =& \frac{-\lambda cn}{\log^{11} n} \left( \frac{1}{n} \right)^{(1+\gamma)(\log 2)^2(1 - \epsilon)/(\lambda \epsilon)}
\end{align*}
to go to 0. Only the highest order terms are relevant, so we can just look at the exponents of $n$, and the expression will go to 0 if
\begin{align*}
    & 1 < \frac{(1+\gamma)(\log 2)^2(1 - \epsilon)}{\epsilon \lambda} \\
    \implies& \epsilon \lambda < (1+\gamma)(\log 2)^2(1 - \epsilon) \\
    \implies& \epsilon \lambda < (1+\gamma)(\log 2)^2 - \epsilon (1+\gamma)(\log 2)^2 \\
    \implies& \epsilon (\lambda + (1+\gamma)(\log 2)^2) < (1+\gamma)(\log 2)^2 \\
    \implies& \epsilon < \frac{(1+\gamma) (\log 2)^2}{\lambda + (1+\gamma) (\log 2)^2}.
\end{align*}

We set $T = (1- \epsilon)n$, so to have success probability going to 1 we must have
\begin{equation*}
    1 - \epsilon > 1 - \frac{(1+\gamma) (\log 2)^2}{\lambda + (1+\gamma) (\log 2)^2} = \frac{\lambda}{\lambda + (1+\gamma) (\log 2)^2},
\end{equation*}
and recall $\gamma$ can be any constant greater than zero.
Altogether this shows the following.
\begin{theorem}
\label{thm:full_lb}
There exists $n_0$ such that for all $n > n_0$, any test scheme using $T$ tests to identify defectives among $n$ items with i.i.d. defective probability $p = \frac{\lambda}{\log n}$ with
\begin{equation*}
    T \leq (1 - \epsilon) \frac{\lambda}{\lambda + \log^2 2} n
\end{equation*}
for some constant $\epsilon$ independent of $n$ must have error probability $1 - o(1)$.
\end{theorem}

\section{Conclusion and Discussion}

We have shown that for nonadaptive PGT in the regime that $p = \lambda / \log n$ (or equivalently $k = \lambda n / \log n$),
\begin{equation*}
    \min \left(1, \left(1+\epsilon\right) \frac{\lambda}{\log^2 2}\right) n
\end{equation*}
measurements suffice to obtain error probability going to 0, and at least
\begin{equation*}
    \left(\left(1 - \epsilon\right) \frac{\lambda}{\lambda + \log^2 2}\right) n
\end{equation*}
are necessary ($\epsilon>0$). \Cref{fig:pgt_bound_comp} shows a graphical comparison of the upper and lower bounds.

\begin{figure}[h!]
\centering
\includegraphics[width=0.7\textwidth]{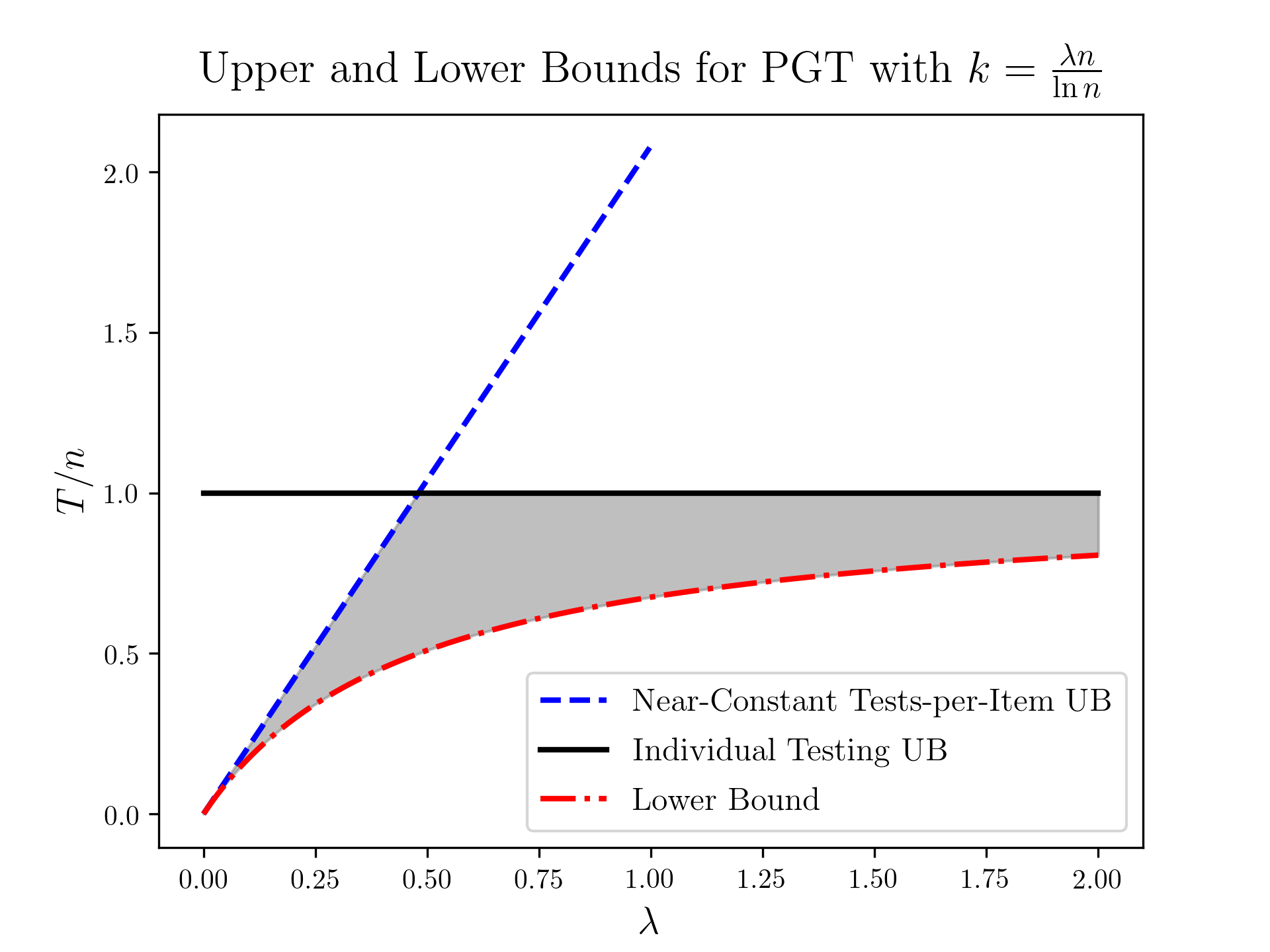}
\caption{Comparison of upper and lower bounds for PGT in the regime where $k = \lambda n / \log n$. The shaded region indicates the gap between the current upper and lower bounds.}
\label{fig:pgt_bound_comp}
\end{figure}

A natural next step would be to close this gap. In the sparse regime where $k = O(n^\theta)$, more complex decodings known as DD and SPIV have been shown to improve on the COMP decoding \cite{johnson2018performance,coja2020optimal}, although the benefit seems to vanish as $\theta$ approaches 1. 

We conjecture that the lower bound should come up to meet the upper bound obtained by the minimum of near-constant tests-per-item designs and individual testing. This would imply that the exact point at which individual testing becomes suboptimal is when $p < (\log 2)^2 / \log n$.

The reason for this belief is as follows. Suppose $D$ is the (random) number of disguised items. A simple calculation shows that,
\begin{equation*}
\E[D] \ge n \exp(T/n \cdot \mathcal{L}_p)
\end{equation*}
where $\mathcal{L}_p$ has been defined in the last section. Using the same calculation as before, we obtain,
\begin{equation*}
\E[D] \ge n 2^{-\frac{T}{n} \frac{\log 2}{p}}.
\end{equation*}
Given the number of disguised items is $D$, the probability of correct decoding is 
at most $(1-p)^D \le \exp(-pD),$ where $p =\frac{\lambda}{\log n}$. Substituting $D$ with $\mathbb{E} D$ from above, we see that the probability of correct decoding is going to zero whenever $\frac{T}{n} < \frac{\lambda}{\log^2 2}.$
This shows that the lower bound should be improved to match the upper bound, if the random variable $D$ is concentrated around its mean, which is of course yet to be proved. %However as of now we are unable to show such concentration

% One possible approach would be to modify the procedure used to construct $W$ to stop with $n - o(n)$ items remaining rather than $\epsilon n$, while managing to keep $|W|$ about $n / \poly(\log n)$.

%%%%%%
%% To balance the columns at the last page of the paper use this
%% command:
%%
%\enlargethispage{-1.2cm} 
%%
%% If the balancing should occur in the middle of the references, use
%% the following trigger:
%%
%\IEEEtriggeratref{4}
%%
%% which triggers a \newpage (i.e., new column) just before the given
%% reference number. Note that you need to adapt this if you modify
%% the paper.  The "triggered" command can be changed if desired:
%%
%\IEEEtriggercmd{\enlargethispage{-20cm}}
%%
%%%%%%

%%%%%%
%% References:
%% We recommend the usage of BibTeX:
%%
\bibliographystyle{IEEEtran}
\bibliography{references}

% Generated by IEEEtran.bst, version: 1.14 (2015/08/26)
\begin{thebibliography}{10}
\providecommand{\url}[1]{#1}
\csname url@samestyle\endcsname
\providecommand{\newblock}{\relax}
\providecommand{\bibinfo}[2]{#2}
\providecommand{\BIBentrySTDinterwordspacing}{\spaceskip=0pt\relax}
\providecommand{\BIBentryALTinterwordstretchfactor}{4}
\providecommand{\BIBentryALTinterwordspacing}{\spaceskip=\fontdimen2\font plus
\BIBentryALTinterwordstretchfactor\fontdimen3\font minus
  \fontdimen4\font\relax}
\providecommand{\BIBforeignlanguage}[2]{{%
\expandafter\ifx\csname l@#1\endcsname\relax
\typeout{** WARNING: IEEEtran.bst: No hyphenation pattern has been}%
\typeout{** loaded for the language `#1'. Using the pattern for}%
\typeout{** the default language instead.}%
\else
\language=\csname l@#1\endcsname
\fi
#2}}
\providecommand{\BIBdecl}{\relax}
\BIBdecl

\bibitem{dorfman1943detection}
R.~Dorfman, ``The detection of defective members of large populations,''
  \emph{The Annals of Mathematical Statistics}, vol.~14, no.~4, pp. 436--440,
  1943.

\bibitem{yelin2020evaluation}
I.~Yelin, N.~Aharony, E.~S. Tamar, A.~Argoetti, E.~Messer, D.~Berenbaum,
  E.~Shafran, A.~Kuzli, N.~Gandali, O.~Shkedi \emph{et~al.}, ``Evaluation of
  covid-19 rt-qpcr test in multi sample pools,'' \emph{Clinical Infectious
  Diseases}, vol.~71, no.~16, pp. 2073--2078, 2020.

\bibitem{gollier2020group}
C.~Gollier and O.~Gossner, ``Group testing against covid-19,'' EconPol Policy
  Brief, Tech. Rep., 2020.

\bibitem{eberhardt2020multi}
J.~N. Eberhardt, N.~P. Breuckmann, and C.~S. Eberhardt, ``Multi-stage group
  testing improves efficiency of large-scale covid-19 screening,''
  \emph{Journal of Clinical Virology}, vol. 128, p. 104382, 2020.

\bibitem{du2000combinatorial}
\BIBentryALTinterwordspacing
D.~Du and F.~Hwang, \emph{Combinatorial Group Testing and Its Applications},
  ser. Applied Mathematics.\hskip 1em plus 0.5em minus 0.4em\relax World
  Scientific, 2000. [Online]. Available:
  \url{https://books.google.com/books?id=KW5-CyUUOggC}
\BIBentrySTDinterwordspacing

\bibitem{chan2011non}
C.~L. Chan, P.~H. Che, S.~Jaggi, and V.~Saligrama, ``Non-adaptive probabilistic
  group testing with noisy measurements: Near-optimal bounds with efficient
  algorithms,'' in \emph{2011 49th Annual Allerton Conference on Communication,
  Control, and Computing (Allerton)}.\hskip 1em plus 0.5em minus 0.4em\relax
  IEEE, 2011, pp. 1832--1839.

\bibitem{sebHo1985two}
A.~Seb{\H{o}}, ``On two random search problems,'' \emph{Journal of Statistical
  Planning and Inference}, vol.~11, no.~1, pp. 23--31, 1985.

\bibitem{atia2012boolean}
G.~K. Atia and V.~Saligrama, ``Boolean compressed sensing and noisy group
  testing,'' \emph{Information Theory, IEEE Transactions on}, vol.~58, no.~3,
  pp. 1880--1901, 2012.

\bibitem{mazumdar2015nonadaptive}
A.~Mazumdar, ``Nonadaptive group testing with random set of defectives,''
  \emph{{IEEE} Trans. Information Theory}, vol.~62, no.~12, pp. 7522--7531,
  2016.

\bibitem{barg2017group}
A.~Barg and A.~Mazumdar, ``Group testing schemes from codes and designs,''
  \emph{IEEE Transactions on Information Theory}, vol.~63, no.~11, pp.
  7131--7141, 2017.

\bibitem{inan2019optimality}
H.~A. Inan, P.~Kairouz, M.~Wootters, and A.~{\"O}zg{\"u}r, ``On the optimality
  of the kautz-singleton construction in probabilistic group testing,''
  \emph{IEEE Transactions on Information Theory}, vol.~65, no.~9, pp.
  5592--5603, 2019.

\bibitem{cai2017efficient}
S.~Cai, M.~Jahangoshahi, M.~Bakshi, and S.~Jaggi, ``Efficient algorithms for
  noisy group testing,'' \emph{IEEE Transactions on Information Theory},
  vol.~63, no.~4, pp. 2113--2136, 2017.

\bibitem{vem2017group}
A.~Vem, N.~T. Janakiraman, and K.~R. Narayanan, ``Group testing using
  left-and-right-regular sparse-graph codes,'' \emph{arXiv preprint
  arXiv:1701.07477}, 2017.

\bibitem{lee2019saffron}
K.~Lee, K.~Chandrasekher, R.~Pedarsani, and K.~Ramchandran, ``Saffron: A fast,
  efficient, and robust framework for group testing based on sparse-graph
  codes,'' \emph{IEEE Transactions on Signal Processing}, vol.~67, no.~17, pp.
  4649--4664, 2019.

\bibitem{bondorf2020sublinear}
S.~Bondorf, B.~Chen, J.~Scarlett, H.~Yu, and Y.~Zhao, ``Sublinear-time
  non-adaptive group testing with o (klogn) tests via bit-mixing coding,''
  \emph{IEEE Transactions on Information Theory}, 2020.

\bibitem{inan2020strongly}
H.~A. Inan and A.~Ozgur, ``Strongly explicit and efficiently decodable
  probabilistic group testing,'' in \emph{2020 IEEE International Symposium on
  Information Theory (ISIT)}.\hskip 1em plus 0.5em minus 0.4em\relax IEEE,
  2020, pp. 525--530.

\bibitem{agarwal2018novel}
A.~Agarwal, S.~Jaggi, and A.~Mazumdar, ``Novel impossibility results for
  group-testing,'' in \emph{2018 IEEE International Symposium on Information
  Theory (ISIT)}.\hskip 1em plus 0.5em minus 0.4em\relax IEEE, 2018, pp.
  2579--2583.

\bibitem{aldridge2018individual}
M.~Aldridge, ``Individual testing is optimal for nonadaptive group testing in
  the linear regime,'' \emph{IEEE Transactions on Information Theory}, vol.~65,
  no.~4, pp. 2058--2061, 2018.

\bibitem{scarlett2016limits}
J.~Scarlett and V.~Cevher, ``Limits on support recovery with probabilistic
  models: An information-theoretic framework,'' \emph{IEEE Transactions on
  Information Theory}, vol.~63, no.~1, pp. 593--620, 2016.

\bibitem{aldridge2017capacity}
M.~Aldridge, ``The capacity of bernoulli nonadaptive group testing,''
  \emph{IEEE Transactions on Information Theory}, vol.~63, no.~11, pp.
  7142--7148, 2017.

\bibitem{johnson2018performance}
O.~Johnson, M.~Aldridge, and J.~Scarlett, ``Performance of group testing
  algorithms with near-constant tests per item,'' \emph{IEEE Transactions on
  Information Theory}, vol.~65, no.~2, pp. 707--723, 2018.

\bibitem{coja2020optimal}
A.~Coja-Oghlan, O.~Gebhard, M.~Hahn-Klimroth, and P.~Loick, ``Optimal group
  testing,'' in \emph{Conference on Learning Theory}.\hskip 1em plus 0.5em
  minus 0.4em\relax PMLR, 2020, pp. 1374--1388.

\bibitem{aldridge2019group}
M.~Aldridge, O.~Johnson, and J.~Scarlett, ``Group testing: an information
  theory perspective,'' \emph{arXiv preprint arXiv:1902.06002}, 2019.

\bibitem{bay2020optimal}
W.~H. Bay, E.~Price, and J.~Scarlett, ``Optimal non-adaptive probabilistic
  group testing requires $\theta (\min \{ k \log n, n\}) $ tests,'' \emph{arXiv
  e-prints}, pp. arXiv--2006, 2020.

\end{thebibliography}
%%
%% where we here have assume the existence of the files
%% definitions.bib and bibliofile.bib.
%% BibTeX documentation can be obtained at:
%% http://www.ctan.org/tex-archive/biblio/bibtex/contrib/doc/
%%%%%%

\end{document}